\documentclass[
    ,final            
  ]
  {aipproc}

\usepackage{amssymb}
\usepackage{amsmath}
\usepackage{amsbsy}
\usepackage{latexsym}
\usepackage{array}

\usepackage{graphicx}
\usepackage{epsfig}
\usepackage{graphicx}
\usepackage{graphics}
\usepackage{epsfig}

\layoutstyle{6x9}

\renewcommand\XFMaddressmarkstyle{\alph}

\newcommand{\s}{\sigma}

\newcommand{\bra}{\langle}
\newcommand{\ket}{\rangle}

\newcommand{\be}{\begin{equation}}
\newcommand{\ee}{\end{equation}}
\newcommand{\bea}{\begin{eqnarray*}}
\newcommand{\eea}{\end{eqnarray*}}

\newcommand{\emdash}{\hspace{1pt}---\hspace{1pt}}
\newcommand{\half}{\frac{1}{2}}

\begin{document}

\title{The Radius of the Proton: Size Does Matter}


\author{Jonathan~D.~Carroll}{address={Centre for the Subatomic Structure of
    Matter (CSSM),\\ 
Department of Physics, University of Adelaide, SA 5005, Australia,\\ 
\url{http://www.physics.adelaide.edu.au/cssm}}}

\author{Anthony~W.~Thomas}{address={Centre for the Subatomic Structure of
    Matter (CSSM),\\
Department of Physics, University of Adelaide, SA 5005, Australia,\\ 
\url{http://www.physics.adelaide.edu.au/cssm}}}

\author{Johann~Rafelski}{address={Departments of Physics and Mathematics,
    University of Arizona, Tucson, Arizona, 85721 USA}}

\author{Gerald~A.~Miller}{address={University of Washington, 
    Seattle, WA 98195-1560 USA}}


\begin{abstract}
  The measurement by Pohl {\it et al.}~\cite{Pohl:2010zz} of the
  2S$_{1/2}^{F=1}$ to 2P$_{3/2}^{F=2}$ transition in muonic hydrogen
  and the subsequent analysis has led to a conclusion that the rms charge
  radius of the proton differs from the accepted
  (CODATA~\cite{Mohr:2008fa}) value by approximately 4$\%$, leading to
  a 4.9\,$\s$ discrepancy. We investigate the muonic hydrogen spectrum
  relevant to this transition using bound-state QED with Dirac
  wave-functions and comment on the extent to which the
  perturbation-theory analysis which leads to the above conclusion can
  be confirmed.\par
\end{abstract}

\maketitle

\section{Introduction}

\begin{figure}
\centering \includegraphics[width=0.6\textwidth]{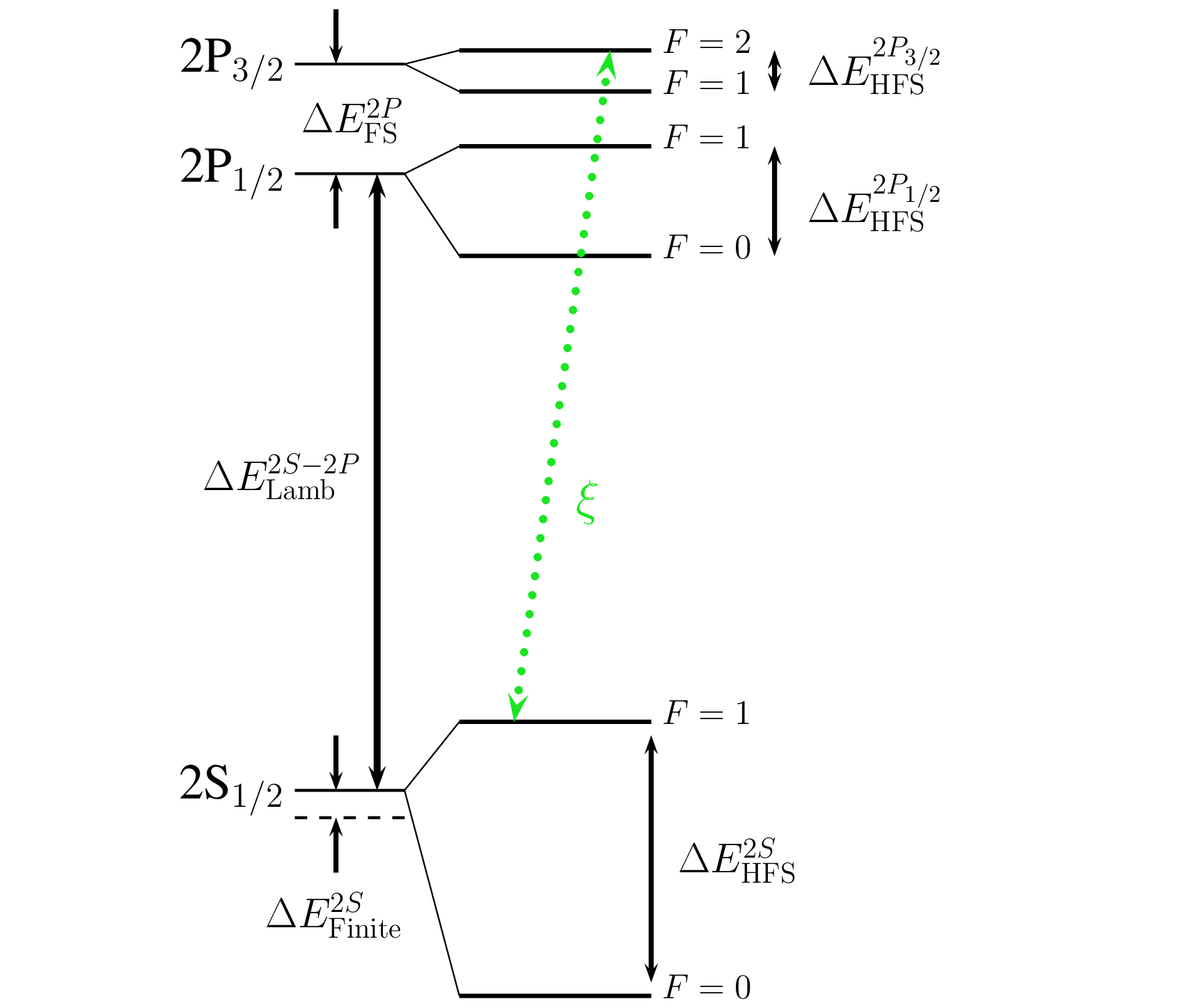}
\caption{(Color online) Muonic hydrogen spectrum, including
  finite-size correction, Lamb shift, fine structure, and hyperfine
  structure. Also shown is the measured 2S$_{1/2}^{F=1}$ to
  2P$_{3/2}^{F=2}$ transition (green, dotted, $\xi$) as per
  Ref.~\cite{Pohl:2010zz}.\protect\label{fig:spectrum} }
\end{figure}

In this work we calculate the transition energy relevant to the
aforementioned experiment of Pohl {\it et al.}~\cite{Pohl:2010zz} (as
depicted in Fig.~\ref{fig:spectrum}) using the Dirac equation in an
attempt to quantify the errors associated with the perturbative
approach. In the sections following, we discuss the nature of the
transition and its components; the method by which we calculate the
energies corresponding to the various eigenstates; and the predicted
energies of the component shifts as a brief account of a longer
upcoming publication~\cite{Carrolldirac} in which we shall detail the
components in full with comparisons to previous
work~\cite{Borie:1982ax,Martynenko:2004bt,Martynenko:2006gz}. We note
that since this talk was presented, we have investigated an additional
term that to our knowledge does not already appear in the analysis of
Pohl~{\it et al.} and which may account for all or part of the
discrepancy~\cite{Miller:2011yw}.


%

\section{Numerical Method}\label{sec:numerical method}

To calculate the theoretical shift corresponding to the measured
transition, previous authors have primarily used perturbation theory
with non-relativistic wave-functions to predict the size of the
contributing effects, including relativistic effects. To better
approximate the exact energies, we can use the Dirac equation for the
muon with the appropriate potential as an effective approximation to
the two-particle Bethe-Saltpeter equation~\cite{Eides:2000xc} to
calculate the perturbed wave-functions, expressed here as a spinor
\be
\psi_\alpha(\vec{r\, }) = \begin{pmatrix}
    g_\alpha(r) \chi_{\kappa}^\mu(\hat{r}) \\[2mm]
    -if_\alpha(r) \chi_{-\kappa}^\mu(\hat{r})
  \end{pmatrix}  = \begin{pmatrix}
    {\displaystyle \frac{G_\alpha(r)}{r}}\ \chi_{\kappa}^\mu(\hat{r}) \\[2mm]
    {\displaystyle \frac{-iF_\alpha(r)}{r}}\ \chi_{-\kappa}^\mu(\hat{r})
  \end{pmatrix},
\ee
normalised to unity, such that the probability is 
\be
\int |\psi_\alpha|^2\; d^3r = \int_0^\infty r^2 \left[g_\alpha(r)^2+f_\alpha(r)^2
\right]\; dr = 1,
\ee
noting that 
\be
\int \chi_\kappa^{m\dagger}\chi_{\kappa'}^{m'}\;
d\hat{r} = \delta_{\kappa\kappa'}\delta_{mm'}.
\ee
Since this is a relativistic system, we use the reduced mass $\mu$ in
place of the muon mass in the Dirac equation
\be
\mu = \frac{M_pm_\mu}{M_p+m_\mu},
\ee
which along with the addition of recoil corrections provides a good
approximation to the Bethe-Saltpeter equation. Since the binding of
the muon in this system is extremely weak, the eigenvalue
$\epsilon_\alpha$ for each state calculated using the Dirac equation
is approximately equal to the reduced mass $\mu$. In order to
precisely calculate the variance from this value, we shift our
eigenvalue down by the reduced mass, such that the eigenvalue we are
now solving for is $\lambda_\alpha = \epsilon_\alpha - \mu$, thus the
effective Dirac equation is
%
\be
\label{eq:dirac}
\frac{d}{dr}\begin{pmatrix}G_\alpha(r)\\[2mm] F_\alpha(r)\end{pmatrix} = \left(\begin{array}{cc}
-{\displaystyle \frac{\kappa_\alpha}{r}} & \lambda_\alpha + 2\mu - V(r) \\[2mm]
-\lambda_\alpha + V(r) & {\displaystyle \frac{\kappa_\alpha}{r}}
\end{array}\right)
\begin{pmatrix}G_\alpha(r)\\[2mm] F_\alpha(r)\end{pmatrix},
\ee
%
where the value of $\kappa_\alpha$ is specific to each eigenstate, namely
\begin{eqnarray*}
{\rm 1S}_{1/2}: \kappa =& -1,\quad {\rm 2S}_{1/2}: \kappa =& -1, \\
{\rm 2P}_{1/2}: \kappa =& +1,\quad {\rm 2P}_{3/2}: \kappa =& -2.
\end{eqnarray*}

The (shifted) eigenvalues can be reliably reproduced by using the
point-Coulomb potential
\be V(r) = -\frac{Z\alpha}{r}, \ee
in Eq.~(\ref{eq:dirac}). In order to integrate Eq.~(\ref{eq:dirac}),
we supply an initial guess for the eigenvalue $\lambda_\alpha$, and
appropriate boundary behaviour of upper and lower components of the
wave-function at small and large radii, then integrate from each limit
towards a central match-point. The discontinuity in the wave-function
integrated from each limit is used as a measure of the inaccuracy of
the eigenvalue and a refined estimate is calculated. This process is
iterated until the change in $\lambda_\alpha$ is less than the
required tolerance, at which point we regard the wave-function to be
converged.

To convince ourselves that our method is self-consistently accurate,
we check the accuracy of our procedure using several methods. The
unperturbed Dirac eigenvalues are known
analytically~\cite{Rafelski:1977vq} to be
\be 
\label{eq:exactDirac}
\lambda_\alpha = \epsilon_\alpha - \mu =
\mu\left[1+\frac{Z^2\alpha^2}{\left(n_\alpha-|\kappa_\alpha|+
\sqrt{\kappa_\alpha^2-Z^2\alpha^2}\right)^2}\right]^{-\half}-\mu, 
\ee
where $n_\alpha$ is the principle quantum number for the state
$\alpha$. We first ensure that we are able to reproduce these
values. For the 2S$_{1/2}$ wave-function, we reproduce this value to
within 0.01~$\mu$eV using quad-precision Fortran, a sufficiently large
grid size, and sufficiently small grid spacing, within reasonable
compute-time.
We also check the validity of the virial theorem for our solutions (refer
to Ref.~\cite{Rafelski:1977vq} for further details) by calculating the
reduced eigenvalue as
\be 
\label{eq:virial}
\lambda = \bra {\rm 2S}_{1/2}\, |\, \mu\beta + V(\vec{r\, }) +
\vec{r\, }\cdot \vec{\nabla}V(\vec{\, r})\, |\, {\rm 2S}_{1/2}\ket -\mu,
\ee
which tests the accuracy of the wave-function at the origin where
$|\vec{\nabla}V|$ is greatest. We calculate that the values obtained
using Eqs.~(\ref{eq:exactDirac}) and (\ref{eq:virial}) differ by
0.18~$\mu$eV for a point-Coulomb potential, and 0.45~$\mu$eV a
finite-Coulomb potential (to be discussed later). We therefore
conservatively take our errors to be of the order of $\sim \pm
0.5$~$\mu$eV.

\section{Numerical Calculations}\label{sec:calcs}

\par

\vspace{3mm}
{\bf \quad $\bullet$ 2S$_{1/2}$--2P$_{1/2}$ Lamb Shift:} The Lamb shift is the
splitting of the otherwise degenerate 2S$_{1/2}$ and 2P$_{1/2}$
eigenstates attributed to the vacuum polarization potential $V_{\rm
  VP}$ 
\be
\label{eq:VP}
V_{\rm VP}(r) =
-\frac{Z\alpha}{r}\frac{\alpha}{3\pi}\int_4^{\infty}\frac{e^{-m_{e}qr}}{q^2}\ 
\sqrt{1-\frac{4}{q^2}}\left(1+\frac{2}{q^2}\right)d(q^2).
\ee
We can calculate the shift in eigenvalues using converged Dirac
wave-functions in response to the Coulomb and vacuum polarization
potentials, and in this case we simply take the difference between the
converged eigenvalues for the 2S$_{1/2}$ and 2P$_{1/2}$ states
\be
\Delta E_{\rm Lamb}^{\rm 2S-2P} = \lambda_{\rm 2P_{1/2}} -
\lambda_{\rm 2S_{1/2}} = 205.1822~{\rm meV}.
\ee
Care must be taken when comparing this calculation to that of
perturbative results since our calculation includes relativistic
corrections, which are included later as corrections in perturbative
calculations, e.g. Ref.~\cite{Borie:2004fv}.

\vspace{3mm}
{\bf $\bullet$ Proton Finite-Size Corrections:} To calculate this
effect in our fully relativistic calculation, we consider the
replacement of the point-Coulomb potential with the finite-size
Coulomb potential in Eq.~(\ref{eq:dirac})
\be \label{eq:finiteC}
V_{C}(r) = -\frac{Z\alpha}{r} \to
-Z\alpha\int\frac{\rho(r')}{|\vec{r}-\vec{r\, }'|}d^3r,
\ee
where $\rho(r)$ is the proton charge-distribution (or more
accurately, the slope of the electric form-factor). We have studied
the dependence of the finite-size correction on the form of this term
(always normalised to unity) and this will be summarized in an
upcoming publication (Ref.~\cite{Carrollformfactor}), though the
dependence on the choice of charge-distribution\emdash whether it be
exponential, Yukawa, or Gaussian in form\emdash appears to be
small. Similarly the finite vacuum polarization potential is given by
a convolution of Eq.~(\ref{eq:VP}) with the charge-distribution.

The exponential form for the charge-distribution, normalised to unity
such that ${\int \rho(r)\, d^3r = 1}$ is given by
\be \label{eq:CD}
\rho(r) = \frac{\eta}{8\pi} e^{-\eta r}; \quad \eta = \sqrt{12/\bra r_p^2\ket}.
\ee

We calculate the Lamb shift by taking the difference between the
appropriate eigenvalues calculated using the Dirac equation with the
potential given by Eq.~(\ref{eq:finiteC}) with the charge-distribution
given by Eq.~(\ref{eq:CD}) for various values of $\bra r_p^2\ket$. We
  then interpolate the energy shifts and fit the data to a cubic of
  the form
\be \label{eq:cubic}
f(x) = A + B \bra r_p^2\ket + C \bra r_p^2\ket^{3/2},
\ee 
which provides the relevant parameterization. The dependence of the
Lamb shift on the rms charge radius in the presence of an exponential
finite-sized Coulomb potential and finite vacuum polarization
potential is given by
\be
\Delta E_{\rm finite} = 205.1822 - 5.2519 \bra r_p^2\ket + 0.0546 \bra r_p^2\ket^{3/2}~{\rm meV}.
\ee

\vspace{3mm}
{\bf $\bullet$ 2P Fine Structure:} Subtracting the converged
eigenvalues of the 2P$_{1/2}$ and 2P$_{3/2}$ eigenstates gives the
fine structure splitting
\be
\Delta E_{FS}^{2P} = \lambda_{2P_{3/2}} - \lambda_{2P_{1/2}}
\ee
which we can also calculate in the presence of the various
potentials. For the case of an exponential finite-Coulomb potential
with finite vacuum polarization potential, the 2P fine structure shift
is
\be
\Delta E_{FS}^{\rm 2P} = 8.4206(5)~{\rm meV}.
\ee
 
The shifts due to finite-size effects (as compared to the point cases)
are below the level of errors for our calculation. The point vacuum
polarization itself increases the fine structure shift by $5~\mu$eV.

\vspace{3mm} 
{\bf $\bullet$ 2S$_{1/2}$ Hyperfine Structure: } The
splitting between the 2S $F=0$ and $F=1$ hyperfine eigenvalues is
given~\cite{weissbluth1978atoms} by
\be \label{eq:2shfs_weisskopf}
\Delta E_{HFS}^{2S(F=1-F=0)} =
\frac{16\pi}{3}\beta\gamma|\psi(0)|^2.
\ee

The value of the 2S hyperfine shift, as calculated using
Eq.~(\ref{eq:2shfs_weisskopf}) with the wave-function calculated with
the Dirac equation in the presence of the combined point-Coulomb and
point vacuum polarization potentials is
\be
\Delta E_{HFS}^{2S} = 22.8967(5)~{\rm meV}.
\ee
 The finite-size effects will be investigated in an upcoming
publication~\cite{Carrolldirac}.

\vspace{3mm}
{\bf $\bullet$ 2P$_{1/2}$ Hyperfine Structure: } The 2P$_{1/2}$
Hyperfine structure is of no consequence to the measured transition we
are investigating. Nonetheless, we calculate the energy of the
2P$_{1/2}^{F=0}$ and 2P$_{1/2}^{F=1}$ states as a confirmation of our
method, and to compare to perturbative results. The 2P hyperfine
structure is given~\cite{weissbluth1978atoms} by
\be 
E_{\rm HFS}^{{\rm 2P}} =
2\beta\gamma\frac{\ell(\ell+1)}{j(j+1)}\bigg\bra\frac{1}{r^3}\bigg\ket
\bra Fm_F\, |\, \mathbf{I}\cdot\mathbf{J}\, |\, Fm_F \ket, 
\ee
where the non-zero terms in the dot-product are given by
\be \label{eq:idotj}
\bra Fm_F\, |\, \mathbf{I}\cdot\mathbf{J}\, |\, Fm_F \ket = \half [F(F+1)-I(I+1)-j(j+1)],
\ee
which, for Schr\"odinger wave-functions gives
\be
\label{eq:2p32weissbluth}
\Delta E_{\rm HFS}^{{\rm 2P}_{1/2}} =
\frac{2}{9}\beta\gamma/a_0^3\ ,
\ee
to which anomalous magnetic moments provide further corrections. Using
the converged Dirac wave-functions with exponential finite-Coulomb and
finite vacuum polarization potentials (rather than Schr\"odinger
wave-functions) we calculate the expectation value of $r^{-3}$ and
find
\be
\Delta E_{\rm HFS}^{{\rm 2P}_{1/2}} = 7.6204(5)~{\rm meV}.
\ee
The addition of the (point) vacuum polarization potential to the
point-Coulomb potential increases the splitting by 0.0017(5)~meV, and
the introduction of the finite-Coulomb potential increases this
further by 0.0045(5)~meV. The finite vacuum polarization potential
does not alter the result from the point case here.

\vspace{3mm} 
{\bf $\bullet$ 2P$_{3/2}$ Hyperfine Structure: }
Following the same method as the 2P$_{1/2}$ calculation, we can
calculate the energy levels for the 2P$_{3/2}^{F=1}$ and
2P$_{3/2}^{F=2}$ eigenstates. Using the converged Dirac wave-functions
we find
\be
\Delta E_{\rm HFS}^{{\rm 2P}_{3/2}} = 3.0415(5)~{\rm meV}
\ee
when the potential consists of the exponential finite-Coulomb and
finite vacuum polarization potentials. For this state, the addition of
the (point) vacuum polarization potential to the point-Coulomb
potential increases the splitting by 0.0007(5)~meV, and the
introduction of the finite-size effects was found to make no change
within the limits of our calculation.

\section{Conclusions}

We find that the Dirac calculations performed here agree well with
perturbative results once appropriate corrections are made (taking
care regarding double-counting of effects). The calculations presented
here and discussions of the comparison to perturbative calculations
will be fully detailed in several upcoming
publications~\cite{Carrolldirac, Carrollformfactor}.


\begin{theacknowledgments}

This research was supported in part by the United States Department of
Energy (under which Jefferson Science Associates, LLC, operates
Jefferson Lab) via contract DE-AC05-06OR23177 (JDC, in part); grant
FG02-97ER41014 (GAM); and grant DE-FG02-04ER41318 (JR), and by the
Australian Research Council and the University of Adelaide (JDC,
AWT). GAM and JR gratefully acknowledge the support and hospitality of
the University of Adelaide while the project was undertaken.

\end{theacknowledgments}



\bibliographystyle{aipproc}   

\bibliography{muHrefs}

\end{document}